# THE VARIABILITY OF QUASARS.
# II. FREQUENCY DEPENDENCE

A. DI CLEMENTE, E. GIALLONGO

Osservatorio Astronomico di Roma, via dell'Osservatorio, I-00040 Monteporzio,

Italy

G. NATALI

Istituto di Astrofisica Spaziale C.N.R., C.P. 67, I-00044 Frascati, Italy

D. TRÈVESE

Istituto Astronomico, Università di Roma "La Sapienza" via G.M. Lancisi 29,

I-00161 Roma, Italy

AND F. VAGNETTI

Dipartimento di Fisica, Università di Roma "Tor Vergata", Via della Ricerca

Scientifica, I-00133 Roma, Italy – Email: vagnetti@roma2.infn.it

ABSTRACT

The variability of 30 PG quasars has been observed in the red band during three years. A rest-frame structure function analysis shows an increase of variability with the time interval up to 0.2 mag r.m.s. after two years. A comparison with the IUE data available for PG quasars shows that the variability increases with the rest-frame frequency at each time interval. Evidence for this effect is supported by the analysis of the average variability of other published quasar samples. The effect can completely account for the increase of variability with redshift found in other studies.

*Subject headings*: Galaxies: Quasars: General

# 1. INTRODUCTION

The study of the quasar variability is of primary importance as a tool for constraining the physics of the emission (Rees 1984, Terlevich et al. 1992, Hawkins 1993), and as a selection criterion (Trèvese et al. 1989, Véron & Hawkins 1995). In both cases, it is essential to understand on statistical grounds the dependence of variability on the intrinsic luminosity of the source, on the redshift and the wavelength. For practical reasons, most variability studies with short time sampling are concentrated on the optically violent variables (OVVs), whose luminosity changes by a large amount over short periods of time. However, they represent a small fraction of the entire population. The majority of quasars selected in flux limited samples exhibit, in the optical band, variations of a few tenths of magnitude, with typical time scales of months to several years (see e.g. Usher, Warnock, Green 1983; Pica et al. 1988). In these samples selection effects can be more easily taken into account. Thus, we have observed a subsample of the Palomar-Green (PG) quasars (Schmidt & Green 1983), which are selected by a uniform and objective ultraviolet excess criterion. Our subsample of 30 objects has been observed with the Schmidt telescope 60/90 cm of the Astronomical Observatory of Rome, equipped with a CCD detector (Natali & Pedichini 1990), which provides a field of $17' \times 25'$, where several bright objects can be used for comparison. Photometry has been obtained in the $R$ band, to optimize the efficiency, with sampling intervals ranging from about one month to 3 years. A statistical comparison with archive IUE data has also been performed to explore the dependence of the variability on the wavelength. Such a dependence has been investigated by Cutri et al. (1985) who directly compared $UBVRIJHK$ observations of a sample of 7 quasars, finding some evidence of a hardening of the spectrum during the bright phase. Such a behaviour has also been found by Kinney et al. 1991 from IUE observation of a subsample of PG quasars. Giallongo, Trèvese and Vagnetti (1991, paper I) have suggested that this behaviour may be responsible for the correlation of the amplitude of variability with redshift, found in their analysis. In this paper we provide a new statistical evidence for the increase of quasar variability with the rest-frame frequency.

## 2. OBSERVATIONS AND DATA ANALYSIS

The 60/90 cm Schmidt telescope of the Astronomical Observatory of Rome is located at Campo Imperatore, on the Apennine mountains, at 2200 m above sea level. The focal length is 180 cm and the pixel size of the 384×576 Thompson 7882, UV coated, CCD chip is $23 \times 23$ $\mu$m. This provides a scale of 2.64 arcsec/pixel.

Since the limiting magnitude of the PG sample is $B \sim 16.5$, in each CCD field of about $17' \times 25'$ centered on a QSO there are 20-40 stars within one magnitude from the QSO, which also appears star-like. This allows a good frame-to-frame calibration and a precise evaluation of the photometric accuracy. Short individual exposures of about 10 minutes were adopted for the brightest quasars to avoid saturation. The resulting total exposure times range from 10 to 20 minutes. Each object was observed three times on average. The sample is presented in Table 1, marked as CIMP (Campo Imperatore). Twilight flat fields were taken for each observing run.

A catalog of reference objects in each quasar field was built up using a simple automatic detection algorithm. The object positions were refined computing the baricenter of the light distribution. At each epoch, relative $R$ magnitudes were computed for three different apertures with radii ranging from 3 to 7 pixels.

For each target QSO a multiple catalog was then derived, which contains the photometry at all the epochs in the three apertures. Magnitude offsets respect to a reference epoch, which was usually the oldest, were computed excluding the QSO itself and those objects with magnitude difference $\geq 3\sigma$, which are less than 10% in the worst case. The average noise, associated with each pair of frames, has been defined as the standard deviation $\sigma_{ij}$ of the magnitude differences of the reference objects in the frames $i$ and $j$. The optimal aperture for each pair of frames was selected as the one giving the lowest noise level. The fraction of light lost does not affect the relative photometry, since it is the same for all the stellar images, including the quasars. The distribution of noise estimates in the optimal apertures is shown in Fig. 1. The modal value, which is representative of the noise standard deviation (see Kendall & Stewart 1963, eq. 11.41), is $\sigma_R \simeq 0.035$. The resulting intrinsic variations $\delta m$ of the QSO magnitude were computed for each target and for each time interval $\Delta t_{rest} = \Delta t/(1+z)$ (see below) and are shown in Fig. 2 and reported in Table 2, where columns 1 and 2 indicate the $\alpha$ and $\delta$ coordinates, column 3 the redshift, column 4 the $B$ magnitude from Schmidt and Green (1993), column 5 the observing time as day, month, year, column 6 the magnitude variation respect to the reference epoch, and column 7 the corresponding standard deviation $\sigma_{ij}$.

## 3. VARIABILITY ANALYSIS

### a) Optical Data

Rather than studying in detail the light curves of a few individual objects, it is possible to derive some general constraints on the emission mechanisms from the ensemble properties of statistical samples. The ensemble statistics is described by the "structure function" which is defined in different ways in the literature. Here we define the structure function as

$$s_p(\Delta t) = \langle |m(t+\Delta t) - m(t)|^p \rangle \times C_p \quad , \tag{1}$$

where the angular brackets indicate the ensemble average, $\Delta t$ is the time interval between two observations, the exponent $p$ may take different values, and $C_p$ is a suitable normalization factor introduced to measure the structure function in units of standard deviation. For $p = 2$ and $C_p = 1$, eq. (1) corresponds to the first-order structure function of Simonetti, Cordes & Heeschen (1985). In this case, when the magnitude variations are due to intrinsic variability plus noise, the two contributions are added in quadrature, and the variability structure function becomes $S_2 = s_2 - \sigma_n^2$, where $\sigma_n^2$ is the noise variance. For $p = 1$, the average is more robust, because it is less sensitive to atypical data. Under the assumption that the statistical distributions of noise and variability have the same shape, the factor $C_1$ has the same value for both, with $C_1 = \sqrt{\pi/2}$ in the case of a gaussian distribution. We then adopt in the following analysis the variability structure function defined as:

$$S_1 = \sqrt{\frac{\pi}{2}\langle |m(t+\Delta t) - m(t)|\rangle^2 - \sigma_n^2}. \tag{2}$$

A similar analysis can be performed using the median instead of the ensemble average, as done by Hook et al. 1994.

If the structure function is computed as a function of the observed time intervals, it depends on the redshift distribution of the quasar sample. Thus it is physically more significant to compute the rest frame structure function after reduction of the time intervals to the rest-frames of individual quasars $\Delta t_r = \Delta t/(1+z)$. The structure function is shown in Fig. 3. The noise $\sigma_n$ is computed as the rms value of the standard deviations $\sigma_{ij}$ associated with the frame pairs $i,j$ contributing to the relevant bins. Each error bar represents the statistical uncertainty on the average of the magnitude differences contributing to the same bin.

At the short time scales $\lesssim 100$ days the variability is only slightly greater than noise, thus it is only marginally significant at the level of accuracy allowed by our photometry. However, a systematic increase corresponding to average variations of about 0.1 mag after 1 year is clearly present.

*b) IUE Data*

To compare our observations of variability in the R band with other wavelengths, we have examined the IUE archive through the European Space Information System (ESIS). We have extracted the available short-wavelength and long-wavelength spectra of the PG objects, excluding: (i) objects classified as Seyfert galaxies ($M_B > -23.5$), (ii) spectra with low S/N ($\lesssim 1$, per pixel), (iii) single epoch observations. The resulting sample consists of the 21 objects marked in Table 1. For about half of these objects, variability in the red is also available. Since the observations of 3C 273 are much more numerous than those of other objects, we have used a randomly extracted subset of observations with comparable time sampling. From both the short and long wavelength spectra we have computed broad band magnitudes in the ranges $1300 - 1900$ Å and $2500 - 3000$ Å, with an arbitrary zero point. From these data we have computed the relevant structure functions (eq. 2), reported in Fig. 3. It appears that at each time interval the amplitude of variability increases with the rest-frame frequency, indicating, on average, a hardening of the spectra in the brighter phases, at least under the assumption that the power-law shape of the spectra is preserved. It should be noted that the latter hypothesis would imply that the structure functions at different wavelengths differ by a multiplicative factor. The presence of different time scales, increasing with wavelength, cannot be excluded by the present data. Notice that a better knowledge of the shape of the structure function in different wavelengths would allow to constrain possible deviations from power-law spectra, even in the absence of simultaneous multifrequency observations.

Our results are consistent with the previous suggestions by Cutri et al. (1985), Kinney et al. (1991) and Paltani & Courvoisier (1994), and i) give a quantitative estimate of the average behavior of the population as a function of time interval; ii) are not biased in favor of highly variable objects, since the objects are extracted from the complete flux limited PG quasars. Of course, in this kind of analysis any phase information is neglected.

## 4. DISCUSSION

The results of our optical observations can be compared with previous studies (e.g Trèvese et al. 1994, Hook et al. 1994, Cristiani, Vio & Andreani 1990), which show larger variability at all time intervals. In general the variability may depend on the intrinsic luminosity, on redshift and wavelength. In particular the dependence on the intrinsic luminosity has been recently studied in a large sample by Hook et al. (1994). This dependence has also been found by Trèvese et al. (1994), although in a smaller and intrinsically fainter sample. For this reason, the comparison should be performed using objects in a given range of absolute magnitudes. The dependence of variability on redshift and frequency is less clear and requires further analysis. In fact, in paper I, it has been found a positive correlation of the amplitude of variability with redshift and it has been suggested the following explanation in terms of spectral variability. A fixed observing frequency corresponds to higher rest-frame frequencies at higher redshifts, thus a hardening of the spectrum in the bright phase, i.e. a larger flux variation at higher frequencies, implies a stronger variability at higher redshifts. The dependence of variability on redshift has been confirmed by Trèvese et al. (1994), and is present also when the analysis is restricted to the objects with $M_B \leq -23.5$. Hook et al. (1994) do not find such correlation in their larger sample. However, the flux limit produces strong correlation between absolute magnitude and redshift, implying that high-redshift objects are intrinsically brighter on average. Thus the intrinsic negative correlation of variability with brightness could mask the positive correlation with redshift.

Simultaneous multiwavelength observations of complete QSO samples, with a uniform coverage of the luminosity–redshift plane, are required to disentangle the various dependencies (see Cristiani et al. 1995). However, also non-simultaneous observations of homogeneous samples at different wavelengths can provide new information about the behaviour of variability.

The dependence of the structure function on frequency, which we have found for the PG sample (see Fig. 3), allows us to elucidate the dependence of variability on frequency and redshift, through a comparison with the optical data from previous studies. We have chosen two reference intervals of $\Delta t$, $0.3 \pm 0.09$ yr and $2 \pm 0.6$ yr, motivated by the fact that at shorter time intervals the estimate of the variability is more affected by the noise, while at longer time intervals high redshift objects are progressively lost, due to the finite time baseline. We have then plotted in Fig. 4 the variability as a function of the average rest frame frequency of each sample. The points in Fig. 4 show an increase of variability with the rest frame frequency at both time intervals. Other existing variability data are derived from blue band samples (Cristiani et al. 1990, Hook et al. 1994, Trèvese et al. 1994) which we have restricted to $z \leq 2.2$ and $M_B < -23.5$ for comparison with our PG data. At the relevant redshifts $\langle z \rangle \sim 1.3 - 1.5$ the average rest-frame wavelengths lie in the interval 1900-1800 Å, close to the rest-frame wavelength of the IUE-long camera for PG quasars which are on average at lower redshift, $\langle z \rangle \simeq 0.3$. The average variability of the above samples for the two adopted time intervals is shown in Fig. 4 and appears consistent with the overall trend despite the different distributions of absolute magnitudes and redshifts. Finally, we have also included in the analysis the variability data in the $U$, $B_J$ and $F$ bands of Cimatti, Zamorani & Marano 1993. Due to the limited time baseline ($\sim 3$ yr in the observer frame) the corresponding points are plotted in Fig. 4 only for the shorter rest-frame time interval of 0.3 yr. These points appear consistent with the overall trend. In particular the increase of variability from the F to the U band is confirmed within the sample, and the U point is fairly consistent with the IUE-short data of the PG sample.

The smaller variability in the red could in part be ascribed to a dilution of the quasar light by the non variable emission of the host galaxy. We estimate that this effect is negligible in our sample for the following reasons. Indeed Elvis et al. 1994 evaluated the fractional contribution $f = I_G/I_{obs}$ of the host galaxy to the observed flux. From their Fig. 6 the value of $f$ at $\log \nu_{rest} = 14.26$ ($\lambda = 1.65 \mu$m) appears to be $\lesssim 20\%$, for objects with $M_V < -24$ corresponding to our limit $M_B < -23.5$. We extrapolate the correction factor to $\log \nu_{rest} = 14.86$ using their template spectrum for the host galaxy obtaining an average correction $f \lesssim 13\%$. Since the observed flux is $I_{obs} = I_Q + I_G = I_Q + f I_{obs}$, then the intrinsic quasar variability $\delta I_Q/I_Q = (1-f)^{-1} \delta I_{obs}/I_{obs}$ increases only by a factor 1.15. According to the above template spectrum, the correction in the UV is smaller and can be neglected. However, even applying the correction only in the red band, the average slope of $S_1$ vs $\log \nu_{rest}$ (Fig. 4) is not appreciably affected.

The increase of quasar variability with rest-frame frequency implies an increase of variability with redshift for samples monitored at a fixed photometric band. This effect has been pointed out by Giallongo et al. (Paper I), who invoked it to explain the positive correlation of variability with redshift found in a composite quasar sample. The slope of a straight line representing data in Fig. 4 is about $\Delta S_1(\Delta t)/\Delta \log \nu \simeq 0.28$ both for $\Delta t = 0.3$ yr and $\Delta t = 2$ yr. This corresponds to the same slope $\Delta S_1(\Delta t)/\Delta \log(1+z) \simeq 0.28$ in the variability-redshift relation, assuming that, during the brightness change, the power-law shape of the individual QSO spectra $\log F_\nu = \alpha \log[(1+z)\nu_{obs}] + const$ is preserved.

In a recent paper, Cristiani et al. (1995) have performed an analysis of the variability of a composite sample, consisting of data from Cristiani et al. (1990), Hook et al. (1994) and Trèvese et al. (1994). The authors confirm the increase of variability with absolute magnitude and the positive correlation of variability with redshift. They show in their Fig. 5 the structure function in intervals of absolute magnitude and redshift. Selecting for comparison the magnitude intervals with $-26.5 \leq \langle M_B \rangle \leq -25.5$, it is possible to estimate $\Delta S_1/\Delta \log(1+z) \simeq 0.25 - 0.30$, in good agreement with our estimate from Fig. 4. Thus, as suggested in Paper I, the entire redshift dependence of variability can be explained in terms of spectral variability.

Recently, gravitational microlensing from a population of intergalactic objects has been studied as a possible cause of quasar variability (Hawkins 1993, 1995, Schneider 1993, Alexander 1995). Achromaticity is a characteristic signature of microlensing of stellar images in the case of "local" galactic lenses. However, in the case of intergalactic microlensing the Einstein ring of the lens can be smaller than the source size. In the standard accretion disk model the high and low energy photons are emitted from the inner and outer parts of the disk, respectively, and this results in a color change during the amplification. While it is difficult to distiguish between this phenomenon and the intrinsic color variations (Baganoff & Malkan 1995), a statistical analysis of the variability as a function of wavelength can constrain models of microlensing-induced variability.

Concerning the intrinsic dependence of variability on frequency, Paltani & Courvoisier (1994) give an explanation in the framework of thermal emission models. Assuming that the variability is due to temperature changes, a hardening of the spectrum in the brighter and hotter phase is expected, provided that the turnover in the spectral energy distribution is not at much higher frequency than the observing band. If the temperature is too high, the slope of the spectrum in the observed optical/UV band becomes less dependent on temperature thus less variable. The increase of variability with frequency which we have shown to be present from the red to the UV can constrain the maximum temperature of any thermal emission model.

We remark that if brighter objects are hotter on average and the variability is due to temperature changes they should be less variable. Indeed, assuming for example a spectrum of black body or thermal bremsstrahlung, the spectral turnover of brigther objects is progressively shifted at higher frequencies producing progressively smaller flux changes from $\delta I/I \propto \delta T/T (h\nu/kT)$ for $kT < h\nu$, to $\delta I/I \propto \delta T/T$ for $kT > h\nu$. This could explain the observed negative correlation of the variability with luminosity.

## 5. SUMMARY

We can summarize our results as follows:

1. The variability of 30 PG quasars has been observed in the red band during three years, with typical accuracy of 0.03 mag (Fig. 1), using a Schmidt with a CCD detector.

2. The analysis of the structure function in the rest frame shows that the variability increases from the noise level at a time interval $\Delta t \sim 1$ month to $\sim 0.2$ mag for $\Delta t \sim 2$ yr.

3. Using public IUE data of a sample of 21 PG quasars, with 12 objects in common with our sample, we have compared the variability in the red and in two UV bands as a function of time interval. The variability appears to increase with the rest-frame frequency at each time interval.

4. The average variability of other published quasar samples has also been computed as a function of their average rest-frame frequency. The results support the evidence for an increase of variability with frequency.

5. The positive correlation of variability with redshift found by Giallongo et al. (Paper I) and recently confirmed by Cristiani et al. 1995 can be entirely accounted for by frequency dependence as already suggested in Paper I.

6. Both the correlations of variability with frequency (and hence redshift) and intrinsic luminosity are qualitatively consistent with thermal spectra becoming hotter, and locally harder, in the bright phase.

*Acknowledgements:* We thank the authors of Cimatti et al. (1993), Cristiani et al. (1990), Hook et al. (1994) for providing us the data in digital form, and S. Cristiani for fruitful discussions. IUE data were extracted and partly analyzed through the European Space Information System. This work was supported by CNR and MURST.

TABLE 1
PG Objects Used in this Work

| $\alpha$ | $\delta$ | $z$ | $B$ | CIMP$^a$ | IUE-long$\lambda^b$ | IUE-short$\lambda^c$ |
|---|---|---|---|---|---|---|
| 00 03 25.0 | +15 53 07 | 0.450 | 15.96 | × | | |
| 00 26 38.1 | +12 59 30 | 0.142 | 14.95 | × | × | × |
| 00 44 31.2 | +03 03 34 | 0.624 | 15.97 | | × | |
| 08 04 35.4 | +76 11 32 | 0.100 | 15.15 | × | × | × |
| 08 44 33.9 | +34 56 09 | 0.064 | 14.00 | × | × | × |
| 09 46 46.4 | +30 09 20 | 1.216 | 16.00 | | | × |
| 10 04 45.1 | +13 03 38 | 0.240 | 15.93 | × | | × |
| 10 48 59.4 | −09 02 13 | 0.344 | 16.00 | × | | |
| 11 00 27.4 | +77 15 08 | 0.313 | 15.86 | × | | |
| 11 16 30.1 | +21 35 43 | 0.177 | 15.17 | | | × |
| 11 38 42.4 | +04 03 38 | 1.876 | 16.05 | × | | |
| 11 48 42.6 | +54 54 13 | 0.969 | 15.82 | × | | × |
| 12 02 08.9 | +28 10 54 | 0.165 | 15.02 | | × | × |
| 12 11 44.8 | +14 19 53 | 0.085 | 14.63 | × | × | × |
| 12 16 47.2 | +06 55 19 | 0.334 | 15.68 | × | | |
| 12 22 56.6 | +22 51 49 | 2.046 | 15.49 | × | | |
| 12 26 33.4 | +02 19 42 | 0.158 | 12.86 | × | × | × |
| 13 02 55.8 | −10 17 16 | 0.286 | 15.09 | | × | × |
| 14 11 50.1 | +44 14 12 | 0.089 | 14.99 | | × | × |
| 14 26 33.8 | +01 30 27 | 0.086 | 15.05 | | × | × |
| 14 35 37.5 | −06 45 22 | 0.129 | 15.44 | × | | |
| 15 12 46.9 | +37 01 56 | 0.371 | 15.97 | × | × | × |
| 15 38 00.6 | +47 45 15 | 0.770 | 16.01 | × | | |
| 15 45 31.1 | +21 01 28 | 0.266 | 16.05 | × | | |
| 16 12 08.7 | +26 11 46 | 0.131 | 16.00 | × | | |
| 16 13 36.3 | +65 50 38 | 0.129 | 15.37 | | × | × |
| 16 26 51.5 | +55 29 05 | 0.133 | 16.17 | × | | |
| 16 30 15.5 | +37 44 08 | 1.471 | 15.96 | | | × |
| 16 34 51.7 | +70 37 37 | 1.334 | 14.90 | × | × | |
| 17 00 13.4 | +51 53 37 | 0.292 | 15.43 | × | | × |
| 17 04 03.5 | +60 48 31 | 0.371 | 15.90 | × | | × |
| 17 15 30.7 | +53 31 24 | 1.920 | 16.30 | × | | |
| 17 18 17.7 | +48 07 11 | 1.084 | 15.33 | × | | |
| 21 12 23.6 | +05 55 12 | 0.466 | 15.52 | × | | |
| 22 33 39.8 | +13 28 21 | 0.325 | 16.04 | × | | |
| 22 51 40.4 | +11 20 41 | 0.323 | 16.25 | × | | |
| 23 02 12.0 | +02 55 34 | 1.044 | 16.03 | × | | |
| 23 08 46.5 | +09 51 55 | 0.432 | 16.12 | × | | × |
| 23 44 03.7 | +09 14 05 | 0.677 | 16.08 | × | | |

$^a$ Campo Imperatore $R$ band
$^b$ 2500–3000 Å
$^c$ 1300–1900 Å

TABLE 2
Journal of the Observations

| $\alpha$ | $\delta$ | $z$ | $B$ | Observation Time | Relative $\delta m$ | $\sigma_{ij}$ |
|---|---|---|---|---|---|---|
| 00 03 25.0 | +15 53 07 | 0.450 | 15.96 | 25 07 90 | — | — |
| | | | | 12 09 90 | $-0.007$ | 0.010 |
| | | | | 14 12 92 | $-0.001$ | 0.038 |
| 00 26 38.1 | +12 59 30 | 0.142 | 14.95 | 19 09 90 | — | — |
| | | | | 19 12 90 | $+0.064$ | 0.016 |
| | | | | 15 12 92 | $-0.244$ | 0.011 |
| 08 04 35.4 | +76 11 32 | 0.100 | 15.15 | 19 03 90 | — | — |
| | | | | 20 03 90 | $+0.002$ | 0.027 |
| | | | | 21 01 91 | $-0.023$ | 0.031 |
| | | | | 22 12 92 | $+0.001$ | 0.038 |
| 08 44 33.9 | +34 56 09 | 0.064 | 14.00 | 24 01 90 | $-0.055$ | 0.013 |
| | | | | 19 03 90 | $+0.008$ | 0.018 |
| | | | | 20 03 90 | $+0.043$ | 0.030 |
| | | | | 20 03 90 | $+0.029$ | 0.031 |
| | | | | 21 01 91 | — | — |
| | | | | 26 01 93 | $+0.257$ | 0.018 |
| 10 04 45.1 | +13 03 38 | 0.240 | 15.93 | 31 01 90 | — | — |
| | | | | 06 03 90 | $-0.050$ | 0.020 |
| | | | | 20 03 90 | $-0.036$ | 0.023 |
| | | | | 21 01 91 | $+0.051$ | 0.018 |
| | | | | 26 01 93 | $-0.032$ | 0.018 |
| 10 48 59.4 | $-09$ 02 13 | 0.344 | 16.00 | 25 01 90 | — | — |
| | | | | 31 01 90 | $+0.132$ | 0.054 |
| | | | | 22 03 90 | $+0.214$ | 0.033 |
| | | | | 22 03 90 | $+0.104$ | 0.029 |
| | | | | 19 03 91 | $+0.180$ | 0.014 |
| | | | | 17 03 93 | $+0.546$ | 0.013 |
| 11 00 27.4 | +77 15 08 | 0.313 | 15.86 | 31 01 90 | — | — |
| | | | | 06 03 90 | $-0.029$ | 0.008 |
| | | | | 29 05 90 | $-0.051$ | 0.027 |
| 11 38 42.4 | +04 03 38 | 1.876 | 16.05 | 22 03 90 | — | — |
| | | | | 30 05 90 | $+0.083$ | 0.040 |
| | | | | 19 03 91 | $+0.076$ | 0.024 |
| 11 48 42.6 | +54 54 13 | 0.969 | 15.82 | 01 02 90 | — | — |
| | | | | 06 03 90 | $-0.013$ | 0.040 |
| | | | | 29 05 90 | $+0.063$ | 0.025 |
| | | | | 20 03 91 | $-0.021$ | 0.049 |
| 12 11 44.8 | +14 19 53 | 0.085 | 14.63 | 01 02 90 | — | — |
| | | | | 28 06 90 | $-0.070$ | 0.026 |
| | | | | 09 04 91 | $+0.129$ | 0.012 |
| 12 16 47.2 | +06 55 19 | 0.334 | 15.68 | 01 02 90 | $+0.122$ | 0.034 |
| | | | | 27 06 90 | $+0.121$ | 0.017 |
| | | | | 13 03 91 | — | — |
| 12 22 56.6 | +22 51 49 | 2.046 | 15.49 | 01 02 90 | — | — |
| | | | | 09 04 91 | $+0.480$ | 0.067 |

| $\alpha$ | $\delta$ | $z$ | $B$ | Observation Time | Relative $\delta m$ | $\sigma_{ij}$ |
|---|---|---|---|---|---|---|
| 12 26 33.4 | +02 19 42 | 0.158 | 12.86 | 01 02 90 | — | — |
| | | | | 30 05 90 | −0.080 | 0.017 |
| | | | | 19 03 91 | −0.231 | 0.034 |
| 14 35 37.5 | −06 45 22 | 0.129 | 15.44 | 30 05 90 | — | — |
| | | | | 17 07 90 | +0.089 | 0.051 |
| 15 12 46.9 | +37 01 56 | 0.371 | 15.97 | 23 07 90 | — | — |
| | | | | 09 04 91 | +0.073 | 0.068 |
| 15 38 00.6 | +47 45 15 | 0.770 | 16.01 | 24 07 90 | — | — |
| | | | | 23 08 90 | −0.030 | 0.012 |
| | | | | 15 04 91 | −0.067 | 0.011 |
| 15 45 31.1 | +21 01 28 | 0.266 | 16.05 | 24 07 90 | −0.049 | 0.022 |
| | | | | 24 08 90 | −0.030 | 0.031 |
| | | | | 15 04 91 | — | — |
| 16 12 08.7 | +26 11 46 | 0.131 | 16.00 | 17 07 90 | — | — |
| | | | | 14 09 90 | +0.111 | 0.038 |
| 16 26 51.5 | +55 29 05 | 0.133 | 16.17 | 25 07 90 | — | — |
| | | | | 28 08 90 | +0.104 | 0.024 |
| 16 34 51.7 | +70 37 37 | 1.334 | 14.90 | 25 07 90 | — | — |
| | | | | 28 08 90 | +0.006 | 0.030 |
| 17 00 13.4 | +51 53 37 | 0.292 | 15.43 | 29 05 90 | — | — |
| | | | | 28 06 90 | +0.060 | 0.018 |
| | | | | 13 09 90 | +0.017 | 0.120 |
| | | | | 11 09 91 | +0.053 | 0.022 |
| 17 04 03.5 | +60 48 31 | 0.371 | 15.90 | 28 06 90 | — | — |
| | | | | 24 07 90 | +0.006 | 0.027 |
| | | | | 14 09 90 | +0.012 | 0.025 |
| 17 15 30.7 | +53 31 24 | 1.920 | 16.30 | 29 05 90 | — | — |
| | | | | 28 06 90 | −0.106 | 0.045 |
| | | | | 13 09 90 | +0.025 | 0.027 |
| | | | | 11 09 91 | +0.010 | 0.030 |
| 17 18 17.7 | +48 07 11 | 1.084 | 15.33 | 28 06 90 | — | — |
| | | | | 24 07 90 | +0.017 | 0.016 |
| | | | | 14 09 90 | +0.033 | 0.023 |
| | | | | 11 09 91 | −0.030 | 0.013 |
| 21 12 23.6 | +05 55 12 | 0.466 | 15.52 | 30 05 90 | — | — |
| | | | | 28 06 90 | −0.175 | 0.084 |
| | | | | 26 07 90 | +0.048 | 0.033 |
| | | | | 24 08 90 | +0.029 | 0.030 |
| 22 33 39.8 | +13 28 21 | 0.325 | 16.04 | 18 07 90 | — | — |
| | | | | 24 08 90 | +0.044 | 0.020 |
| | | | | 12 09 90 | +0.045 | 0.024 |
| 22 51 40.4 | +11 20 41 | 0.323 | 16.25 | 17 07 90 | — | — |
| | | | | 14 09 90 | −0.018 | 0.014 |
| 23 02 12.0 | +02 55 34 | 1.044 | 16.03 | 24 07 90 | — | — |
| | | | | 14 09 90 | −0.005 | 0.019 |
| | | | | 19 12 90 | −0.025 | 0.018 |
| 23 08 46.5 | +09 51 55 | 0.432 | 16.12 | 24 07 90 | — | — |
| | | | | 14 09 90 | +0.041 | 0.063 |
| | | | | 19 09 90 | +0.027 | 0.014 |
| 23 44 03.7 | +09 14 05 | 0.677 | 16.08 | 24 07 90 | — | — |
| | | | | 19 09 90 | +0.018 | 0.008 |
| | | | | 19 12 90 | +0.033 | 0.016 |

# FIGURE LEGENDS

Fig. 1 Distribution of photometric errors.

Fig. 2 Individual magnitude variations $|\delta m|$ as a function of the rest frame time interval.

Fig. 3 Structure functions of the PG objects studied in this work. CIMP: Campo Imperatore R band; IUE-long$\lambda$: 2500-3000 Å; IUE-short$\lambda$: 1300-1900 Å. The abscissa of each point is the average time interval of the data in the relevant bin.

Fig. 4 Variability as a function of the rest-frame frequency, for two different time intervals. White symbols: 0.3 yr; filled symbols: 2 yr.

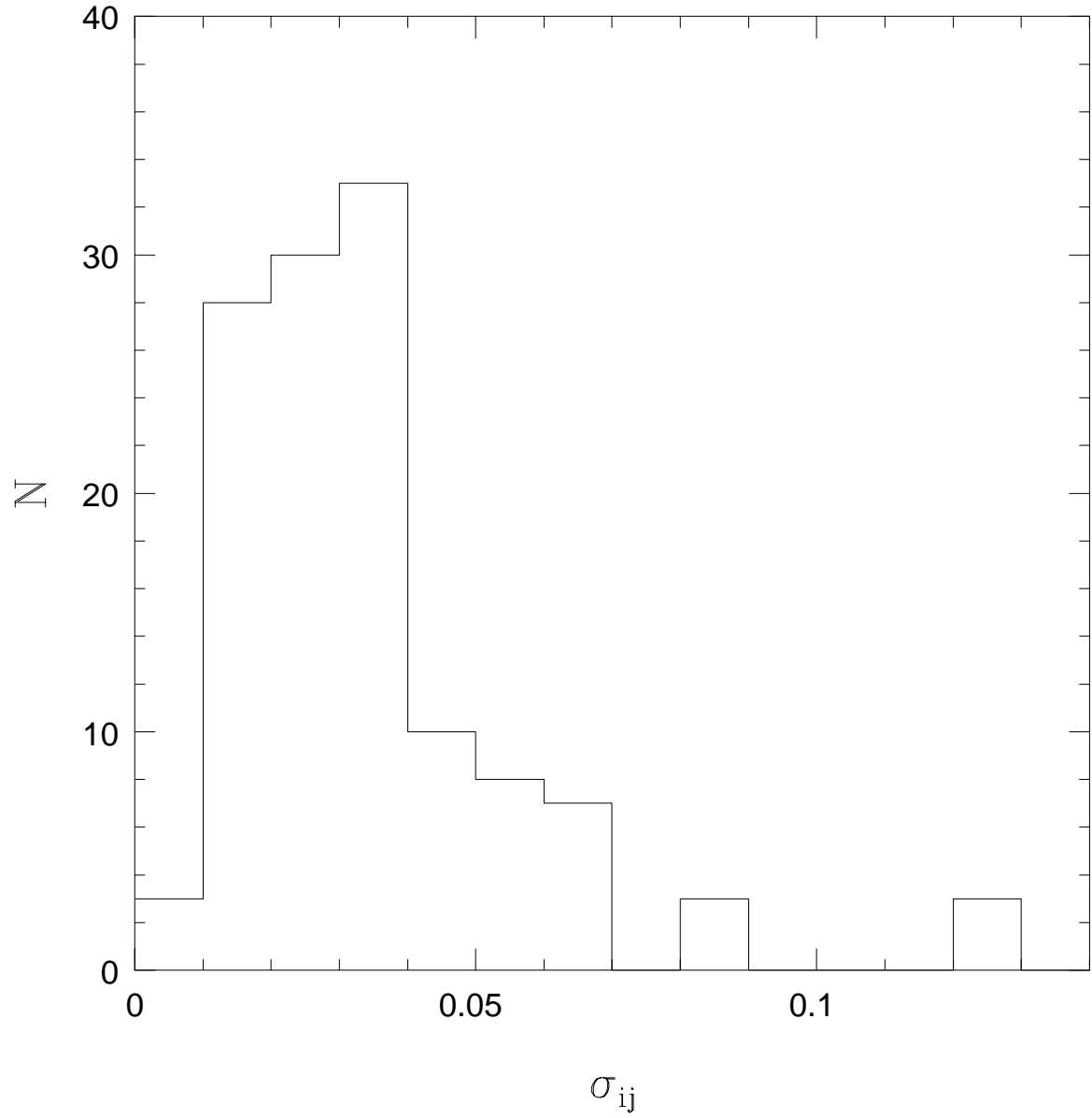

Fig. 1

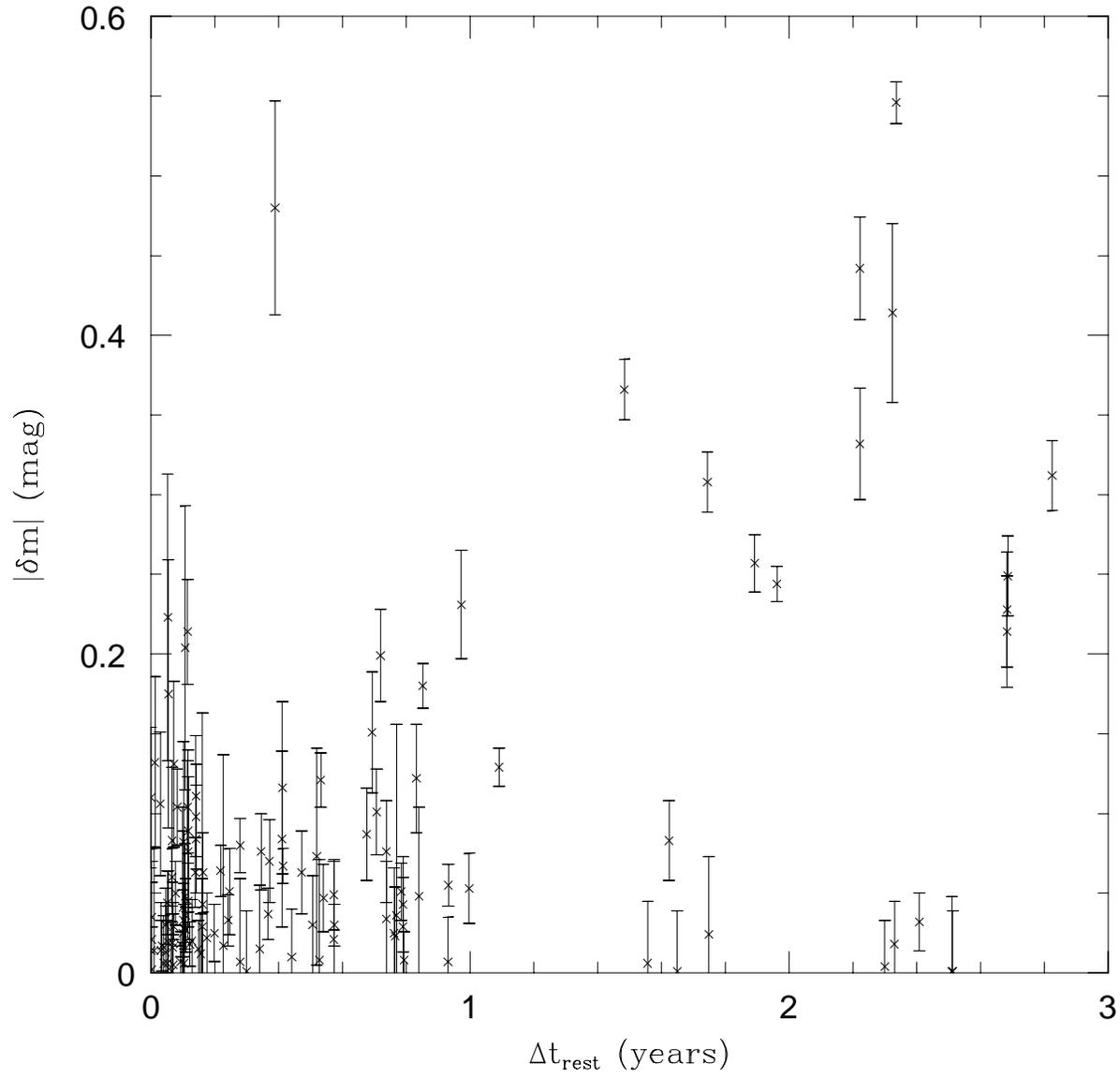

Fig. 2

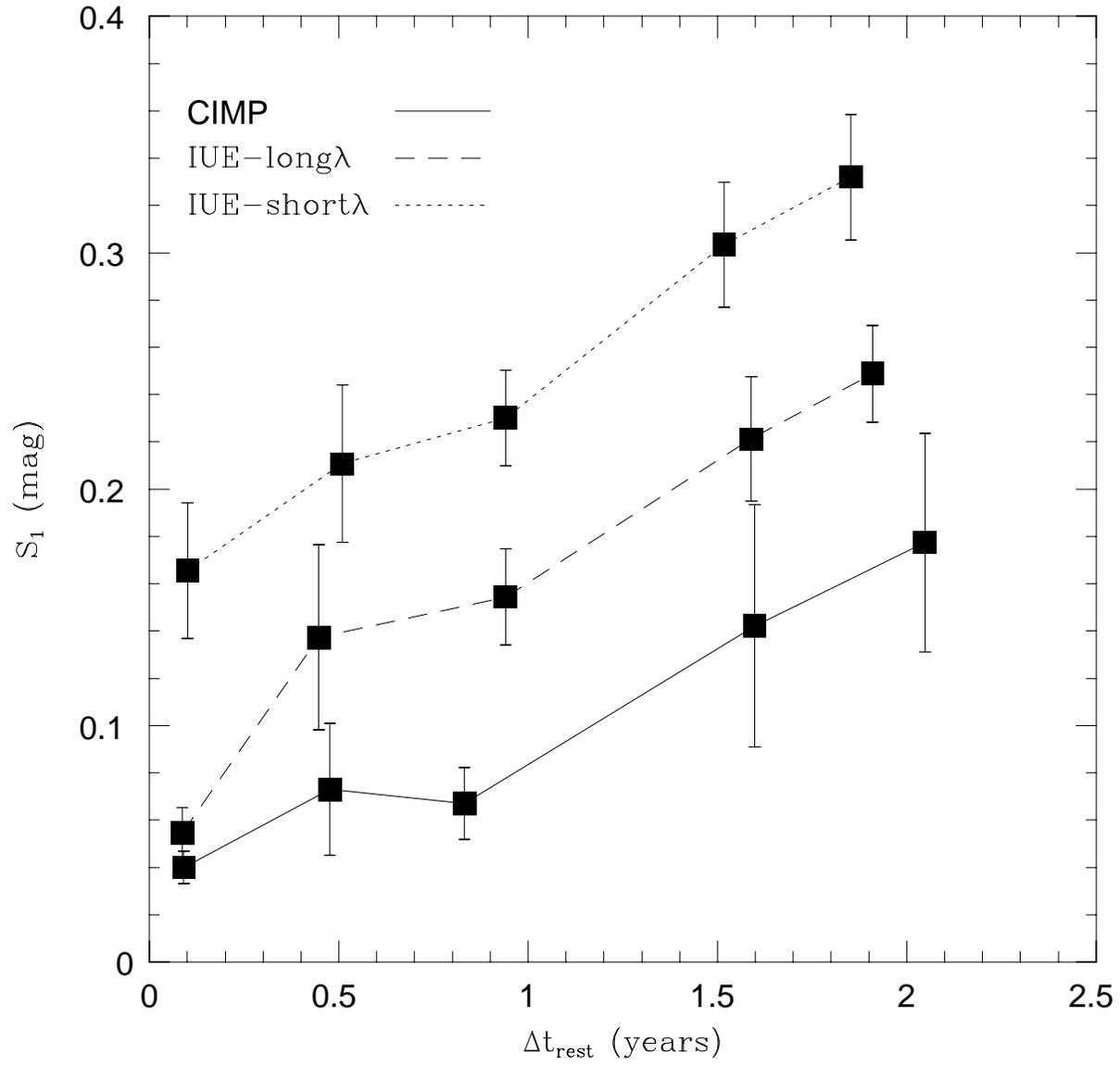

Fig. 3

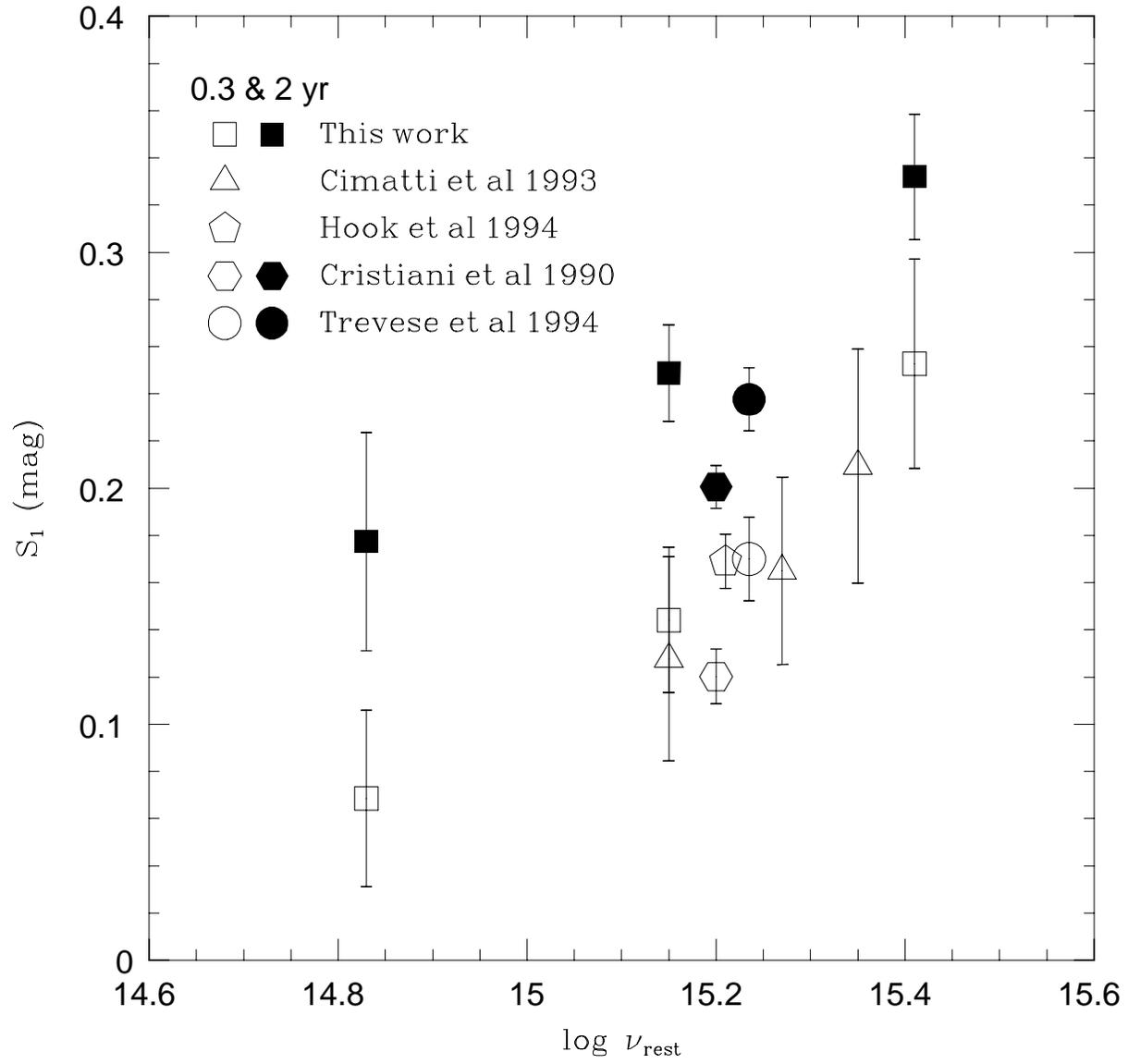

Fig. 4